\documentclass[prd,preprintnumbers,floatfix,aps,nofootinbib,notitlepage]{revtex4}

\usepackage{latexsym}
\usepackage{epsfig}
\usepackage{amssymb}
\usepackage{amsmath}
\usepackage{color}
\usepackage{mathtools}
\usepackage{hhline}

\newcommand{\lp}{\left(}
\newcommand{\rp}{\right)}
\newcommand{\lb}{\left[}
\newcommand{\rb}{\right]}

\newcommand{\ba}{\begin{eqnarray}}
\newcommand{\ea}{\end{eqnarray}}
\newcommand{\be}{\begin{equation}}
\newcommand{\ee}{\end{equation}}

\newcommand{\al}{\alpha}
\newcommand{\bt}{\beta}
\newcommand{\ga}{\gamma}

\newcommand{\la}{\lambda}

\newcommand{\en}{\epsilon}

\newcommand{\Sa}{\Sigma}

\newcommand{\p}{p}

\newcommand{\cM}{\mathcal{M}}
\newcommand{\e}{e^{-2\al+4\sigma}}
\newcommand{\rar}{\rightarrow}

\newcommand{\eN}{{\scriptscriptstyle{\mathcal{N}}}}
\newcommand{\tg}{\tilde{\gamma}}

\newcommand{\lK}{\lambda_K}
\newcommand{\lM}{\lambda_M}
\newcommand{\lV}{\lambda_V}

\newcommand{\SQRT}{\sqrt{\star}}

\begin{document}

\title{Doubly-boosted vector cosmologies from disformal metrics}

\author{Tomi S.~Koivisto}
\email{tomik@astro.uio.no}
\affiliation{Nordita, KTH Royal Institute of Technology and Stockholm University, Roslagstullsbacken 23, SE-10691 Stockholm, Sweden}

\author{Federico R.~Urban}
\email{furban@ulb.ac.be}
\affiliation{Service de Physique Th\'eorique, Universit\'e Libre de Bruxelles, CP225, Boulevard du Triomphe, B-1050 Brussels, Belgium}
\affiliation{Nordita, KTH Royal Institute of Technology and Stockholm University, Roslagstullsbacken 23, SE-10691 Stockholm, Sweden}

\date{\today}

\begin{abstract}

A systematic dynamical system approach is applied to study the cosmology of anisotropic Bianchi I universes in which a vector field is assumed to operate on a disformal frame.  This study yields a number of new fixed points, among which anisotropic scaling solutions.  Within the simplifying assumption of (nearly) constant-slope potentials these are either not stable attractors, do not describe accelerating expansion or else they feature too large anisotropies to be compatible with observations.  Nonetheless, some solutions do have an appeal for cosmological applications in that isotropy is retained due to rapid oscillations of the vector field.

\end{abstract}

\preprint{NORDITA-2015-27}

\maketitle

%%%%%%%%%%%%%%%%%%%%%%%%%%%%%%%%%%%%%%%%%%%%%%%%%%%%%%%%%%%%%%%%%%%
%%%%%%%%%%%%%%%%%%%%%%%%%%%%%%%%%%%%%%%%%%%%%%%%%%%%%%%%%%%%%%%%%%%
%%%%%%%%%%%%%%%%%%%%%%%%%%%%%%%%%%%%%%%%%%%%%%%%%%%%%%%%%%%%%%%%%%%

%%%%%%%%%%%%%%%%%%%%%%%%%%%%%%%%%%%%%%%%%%%%%%%%%%%%%%%%%%%%%%%%%%%
\section{Introduction}
%%%%%%%%%%%%%%%%%%%%%%%%%%%%%%%%%%%%%%%%%%%%%%%%%%%%%%%%%%%%%%%%%%%

The observed large-scale homogeneity and isotropy of the Universe are usually explained within the framework of cosmological inflation.  This epoch of primordial exponential acceleration is traditionally assumed to be driven by a scalar field; one immediate consequence of this choice is that whatever primordial anisotropies might be present at that time, they are rapidly washed out as inflation proceeds, leaving the Universe with an order $10^{-5}$ initial scalar (and somewhat smaller tensor) seeds for the observed structures, germinating from the quantum fluctuactions of the inflaton scalar field itself (see~\cite{Mazumdar:2010sa} for a review).  Scalar fields are also often invoked in models of late-time dark energy, according to which the present acceleration of the Universe is due to the presence of a tiny, smoothly distributed vacuum energy associated with the so called quintessence field (see, e.g.,~\cite{Clifton:2011jh}).

Despite their attractiveness and simplicity however, the existence of these cosmological scalar fields has not yet been established observationally, leaving the window open for other possibilities, among which the logical simplest option is vector fields.  Vector fields in this context could either play the r\^oles of cosmological \emph{prima donnas} as either the actual inflaton or quintessence fields driving the early or the late acceleration, or as supporting acts to a scalar field, which may have non-trivial observable consequences.  For instance, one motivation for entertaining such ideas is the apparent tension among several features in the observed cosmic microwave background radiation and the statistically isotropic standard model of the Universe, see e.g.\ Ref.~\cite{Copi:2013jna,Ade:2013zuv} for recent studies.

Of course, in any reasonable model of our Universe, isotropy (at least approximate) must be had in order to not clash with observations; this means that one has to consider non-minimal options for the vector fields configurations in one way or another.  One could support isotropic expansion long enough with sufficiently finely tuned initial conditions as in the vector field inflation model of~\cite{Ford:1989me}, or invoke a regiment of vector fields in such a way that on average they do not manifest any preferred direction, as in the model put forward by~\cite{Golovnev:2008cf,Golovnev:2008hv}, see also~\cite{Himmetoglu:2008hx,Pitrou:2008gk,Golovnev:2009ks,Golovnev:2011yc,Germani:2009iq}.  With non-minimal coupling to gravity or non-standard vector fields, such as in three-form inflation and dark energy models, a time-like (and thus isotropic) field would include non-trivial (yet observationally appropriate) cosmologies~\cite{Koivisto:2008xf,Jimenez:2009py,Koivisto:2009ew,Koivisto:2009fb,Koivisto:2009sd,Boehmer:2011tp,Ngampitipan:2011se,Koivisto:2012xm,DeFelice:2012wy,DeFelice:2012jt,Koivisto:2011rm,Urban:2012ib,Mulryne:2012ax,Kumar:2014oka}.  Also, one may arrange a ``triad'' of three vector fields in such a way that their vacuum configuration is isotropic~\cite{ArmendarizPicon:2004pm}.  Finally, rapidly oscillating vectors could naturally yield isotropic cosmology~\cite{Cembranos:2012kk}.

Here we investigate a novel class of scalar-vector cosmologies.  In particular we are interested in cosmological dynamics that could yield the so-called ``anisotropic hair'', i.e., vectors which survive the inflationary stretch.  Such scenarios would manifest themselves as attractor solutions in which the isotropic scalar field dominates the energy budget of the Universe, but a small, constant amount of anisotropy is supported by subdominant vector fields.  Such solutions have been found in non-minimally coupled scalar-vector models~\cite{Kanno:2008gn,Watanabe:2009ct,Hervik:2011xm}, see also~\cite{Akarsu:2009gz,Kumar:2011rn,Thorsrud:2012mu,Akarsu:2013dva} for studies of the dynamics of anistropic vector cosmologies.  Here we will study a system in which the non-minimal coupling is of the disformal type: the (generically massive) vector field is thus coupled to the gradients of the scalar field.  We study a simple axisymmetric Bianchi I spacetime and its anisotropic cosmology, in which the direction of the space-like vector determines the preferred axis.

The formalism we employ was recently introduced in Ref.~\cite{Koivisto:2014gia}.  In that case we focussed on a particular extra-dimensional theory in which the vector matter lives on a moving brane~\cite{KOIVISTO:2013jwa,Koivisto:2013fta,Koivisto:2012za}, giving rise to a disformal coupling of the vector to the scalar.  Since in that simplest model we could not realise any ``anisotropic hair'' solution we are led to explore minimal tweaks to it with the hope of finding new more interesting cosmological solutions.  The class of Lagrangians we study here are a ``doubly-boosted'' version of that studied in our previous work, thanks to an additional Lorenz boost factor (due to the brane movement).  The tools for the phase space analysis developed in Ref.~\cite{Koivisto:2014gia}, as anticipated in that work, are extremely handy for the present class of models, and this work is one more proof of their versatility.  After briefly defining our double-boost class in the following section~\ref{set-up}, we formulate the cosmological equations of motions in the language of dynamical systems using the formalism of Ref.~\cite{Koivisto:2014gia}.  The actual analysis is performed in section~\ref{phase}, and we conclude in section~\ref{conclusions}.

%%%%%%%%%%%%%%%%%%%%%%%%%%%%%%%%%%%%%%%%%%%%%%%%%%%%%%%%%%%%%%%%%%%
\section{The action}\label{set-up}
%%%%%%%%%%%%%%%%%%%%%%%%%%%%%%%%%%%%%%%%%%%%%%%%%%%%%%%%%%%%%%%%%%%

The action for this analysis is given by
\be \label{action}
S=\int d^4x \sqrt{-g}\lb \frac{R}{16\pi G} + p(\phi,X)\rb - S_A \,.
\ee
The first piece includes standard Einstein gravity and the scalar with a lagrangian that depends on the field and its kinetic term $X=-g^{\al\bt}\phi_{,\al}\phi_{,\bt}/2$. For a canonic scalar field one has $p=X-V$, where $V(\phi)$ is the scalar potential ($p$ is in fact the ``pressure'' of the field in fluid description).  In our notation a comma stands for partial derivatives.  The second piece is a disformally-inspired Proca lagrangian for the vector field, which explicitly reads
\be
S_A = \int d^4 x \sqrt{-g} \lb \ga^{-2}\lp -\frac14 F^2 + \frac12 M^2A^2\rp + \frac12 K\phi^{,\al}\phi^{,\bt}\lp F_{\ga\al}F^{\ga}_{\phantom{\ga}\bt} 
+M^2A_\al A_\bt \rp \rb \,,
\ee
where $K \equiv K(\phi)$ and $M^2 \equiv M^2(\phi)$.  The field strength tensor is $F_{\al\bt}=A_{\bt,\al}-A_{\al,\bt}$ as usual.  The Lorentz-like factor is given by
\be \label{lorentz}
\ga^{-2} \equiv 1-2 K X \,.
\ee
This action can be obtained from a more compact form realisation of a system in which the vector field lives on a disformal metric of the kind
\be \label{disformal}
\hat{g}_{\mu\nu}=C(\phi)g_{\mu\nu}+D(\phi)\phi_{,\mu}\phi_{,\nu} \,,
\ee
where $C(\phi)$ and $D(\phi)$ are the conformal and disformal factors respectively.  In this work in particular we study a variant of the simpler form which was investigated in \cite{Koivisto:2014gia}.  In general one finds that $K = D/C$.

The next step is to obtain explicit covariant field equations for this system, which are
\be
G_{\mu\nu} = \kappa^2\lp T^A_{\mu\nu} + T^{\phi}_{\mu\nu}\rp\,,
\ee
where the gravitational coupling constant is $\kappa^2 \equiv 1/8\pi G$; the stress energy tensor for the scalar field $\phi$ is
\be \label{step}
T^{\phi}_{\mu\nu} = \p g_{\mu\nu} + \p_{,X}\phi_{,\mu}\phi_{,\nu} \,,
\ee
while that of the vector is
\ba
T^A_{\mu\nu} & = & F_{\mu\ga} F_\nu^{{\phantom{\ga}\ga}} + \lb \ga^{-2}\lp -\frac14 F^2 - \frac12 M^2A^2\rp + \frac12 K\phi^{,\al}\phi^{,\bt}\lp F_{\ga\al}F^{\ga}_{\phantom{\ga}\bt} 
+M^2A_\al A_\bt \rp \rb g_{\mu\nu} \nonumber \\
& + & K \lb \frac12 \phi_{,\mu}\phi_{,\nu} F^2 - \phi^{,\al}\phi^{,\bt} F_{\mu\al} F_{\nu\bt} + \phi_{,\ga}\phi^{,\ga} F_{\mu\al} F_{\nu}^{\phantom{\ga}\al} - \phi_{,\mu}\phi^{,\al} F_{\al\ga} F_{\nu}^{\phantom{\nu}\ga} - \phi_{,\nu}\phi^{,\al} F_{\al\ga} F_{\mu}^{\phantom{\ga}\ga} \rb \nonumber \\
& + & M^2 \lb \ga^{-2} A_\mu A_\nu + K\phi_{,\mu}\phi_{,\nu} A^2 + K\lp\phi_{,\mu} A_\nu+\phi_{,\nu}A_\mu\rp \rb \,.
\ea
Now, since there is an explicit coupling term between the scalar field and the vector, the stress-energy tensors of the two are not separately conserved, but, as it must, $\nabla_\mu(T_{\phi}^{\mu\nu}+T_A^{\mu\nu}) = 0$.

We now set out to study these equations of motiong in a general case, where, in order to allow for the possibility that anisotropies can develop, we have to employ anisotropic spacetimes.

%%%%%%%%%%%%%%%%%%%%%%%%%%%%%%%%%%%%%%%%%%%%%%%%%%%%%%%%%%%%%%%%%%%
\section{Cosmology}\label{cosmology}
%%%%%%%%%%%%%%%%%%%%%%%%%%%%%%%%%%%%%%%%%%%%%%%%%%%%%%%%%%%%%%%%%%%

We choose to work with a Bianchi I axisymmetric metric, since it describes an anisotropic expansion with the anisotropy pointing in one particular direction (we choose the latter to be aligned with the $\hat z$ direction).  The metric is
\be
ds^2 = - dt^2 + e^{2\al(t)} \lb e^{2\sigma(t)} \lp dx^2 + dy^2 \rp + e^{-4\sigma(t)} dz^2 \rb \,.
\ee
Here $\al$ is the overall isotropic volume expansion rate (the Hubble parameter), and $\sigma$ is the anisotropic shear. We can project the covariant equations of motion for this metric with the result:
\ba
0 & = & \ddot{A} + (\dot{\al} + 4\dot{\sigma})\dot{A} + \lp 1 - K\dot{\phi}^2 \rp M^2A \,, \label{first} \\
0 & = & \lp \ddot{\phi} + 3\dot{\al}\dot{\phi} \rp p_{,X} + \ddot{\phi}\dot{\phi}^2 p_{,XX} + p_{,\phi X} \dot{\phi}^2 - p_{,\phi} + 2 K M^2 \e\dot{A}A\dot{\phi} \nonumber \\
& + & \e A^2 \lb K M^2 \lp \ddot{\phi} + (\dot{\alpha} + 4\dot{\sigma})\dot{\phi}\rp + \frac12 K'M^2 \dot{\phi}^2 + MM'\lp 1 + K\dot{\phi}^2 \rp \rb \,, \\
0 & = & 3\ddot{\sigma} + 9\dot{\al}\dot{\sigma} - \e\dot{A}^2 + \e A^2 M^2 \lp 1 - K\dot{\phi}^2 \rp \,, \\
0 & = & 6\ddot{\al} + 9\dot{\al}^2 + 9 \dot{\sigma}^2 + 3 p + \frac12 \e\dot{A}^2 - \frac12 \e A^2 M^2 \lp 1 - K\dot{\phi}^2 \rp \,, \label{fried2}\\
0 & = & 3\dot{\al}^2 - 3\dot{\sigma}^2 - p_{,X}\dot{\phi}^2+ p - \frac12 \e\dot{A}^2 - \frac12 \e A^2 M^2 \lp 1 + K\dot{\phi}^2 \rp\,. \label{last}
\ea
At this point we find it convenient to define the inverse of the Lorentz factor
\be
\tg = 1/\ga = \sqrt{1-K\dot{\phi}^2} \,.
\ee

It is most suitable, for studying the dynamics of the system, to employ the e-folding number $\eN\equiv\al$ as the time-evolution variable.  Also, to be concrete, we choose the scalar field action as the DBI type $$p(X,\phi) = \frac{1-\tg(X,\phi)}{K(\phi)} - V(\phi)\,.$$  After several manipulations we found that the best first-order variables which make the system the most transparent and easy to follow are
\be \label{variables}
x = \frac{1}{\sqrt{3\tg(\tg+1)}}\kappa\phi_{,\eN} \,, \quad y = \sqrt{\frac{\kappa^2V}{3\dot{\al}^2}} \,, \quad
u = e^{-\al+2\sigma}\frac{\sqrt{2-\tg^2}M}{\sqrt{6}\dot{\al}}\kappa A \,, \quad v = \frac{e^{-\al+2\sigma}}{\sqrt{6}}\kappa A_{,\eN} \,, \quad  \cM = \frac{M}{\dot{\al}} \,.
\ee
Notice that $\cM$ simply tells us how $M$ evolves compared to the overall expansion rate, and does not encode any direct dynamic effect upon the background geometry. The derivatives of the metric are described by the shear and the slow-roll parameter
\be
\Sa= \sigma_{,\eN}\,, \quad \en = -\ddot{\al}/\dot{\al}^2 \,.
\ee
The Friedmann equation (\ref{last}) then beautifully reads
\be \label{friedmann}
1-\Sa^2 = x^2+y^2+ u^2 + v^2 \,,
\ee
and the second Friedmann equation (\ref{fried2}) constrains $\en$, almost as beautifully, as
\be
2\en = 3(\Sa^2+1)+3\tg x^2-3y^2- (2\ga^2-1)^{-1}u^2+v^2 \,.
\ee
To close the system we choose to work with the simplest slopes for the three $\phi$-dependent functions, that is, we assume them to be approximately constant, which is equivalent to assuming a nearly exponential form for these quantities.  Therefore, we can define the, approximately time-independent, slope parameters
\be
\la_V = \sqrt{\frac32}\frac{V_{,\phi}}{\kappa V}\,, \quad \la_K = \sqrt{\frac32} \frac{K_{,\phi}}{\kappa K}\,, \quad \lM = 2\sqrt{\frac32} \frac{M_{,\phi}}{\kappa M} \,.
\ee

All ingredients are in place for the dynamical equations, which can be cast as
\ba
2\tg^2 x_{,\eN} & = & (1+2\tg)(1-\tg) \lb \SQRT \lK x - \en \rb x + \Upsilon \,, \\
y_{,\eN} & = & \lp \SQRT \lV x + \en \rp y \,, \\
u_{,\eN} & = & \lp \SQRT \lM x + \en + 2\Sa-1 + \frac{\tg\tg_{,\eN}}{\tg^2-2} \rp u + \sqrt{2-\tg^2} \cM v \,, \\
v_{,\eN} & = & \lp \en - 2\Sa - 2 \rp v - \frac{\tg^2}{\sqrt{2-\tg^2}} \cM u \,, \\
\Sa_{,\eN} & = & \lp \en - 3 \rp \Sa + 2 v^2 + \frac{2\tg^2}{\tg^2-2} u^2 \,, \\
\cM_{,\eN} & = & \lp \SQRT \lM x + \en \rp \cM \,,
\ea
where
\be \label{dlng}
\tg\tg_{,\eN} = \frac{\tg-1}{x} \left\{ (\tg+1) \lb \SQRT \lK x^2 - \en x \rb + \Upsilon \right\} \,;
\ee
notice that we have eliminated $K$ thanks to $3x^2K = \ga-1 = (1-\tg)/\tg$. We have kept aside the second derivative of the scalar field, which only appears in the combination
\ba
\Upsilon & \equiv & \sqrt\frac{\tg+1}{3\tg} \kappa\phi_{,\eN\eN} \\
& = & \frac{1}{2(\tg-1)\tg^2u^2 + (\tg^2-2)x^2} \left\{ \SQRT x^2 \lb \tg^4 \lp 2\lK x^2 -2\lV y^2 -2 (\lK+\lM) u^2 \rp \right.\right. \nonumber\\
&& \left. - \tg^3 \lK x^2 + \tg^2 \lp - 5\lK x^2 + 4\lV y^2 + 2(\lK+2\lM)u^2 \rp + 2\tg\lK x^2 + 2\lK x^2 \rb \nonumber \\
&& \left. - (\tg+1)x \lb (3\tg^2-\en)(\tg^2-2)x^2 + 4\tg^2(\tg-1)\sqrt{2-\tg^2} \cM u v + 2\tg^2(\tg-1)\lp4\Sa+1-\en\rp u^2 \rb \right\} \,; \nonumber
\ea
this combination is manifestly finite in the two extreme limits of $\tg\rar1$ and $\tg\rar0$, which are those we will focus on in the subsequent analysis.  Finally, we use the shorthand $\SQRT \equiv \sqrt{\tg(\tg+1)/2}$.

We can use the Friedmann constraint (\ref{friedmann}) to solve away one of the variables in the system, which we choose to be $y$ assuming it is always positive \footnote{Negative energy density potentials $V<0$ could be allowed by extending the analysis to allow imaginary $y$.}; the phase space is therefore a four dimensional one.  As anticipated above, we study the system in the so-called ultra-relativistic ($\tg=0$) and non-relativistic ($\tg=1$) regimes.

It is interesting to notice a peculiar feature of this system: the coupling of the scalar and the vector is turned off when the mass vanishes $M = 0$, that is, when the vector field is conformally coupled to $\tilde g_{\mu\nu}$ it also decouples altogether from the scalar field $\phi$ in the physical metric $g_{\mu\nu}$.  Nonetheless, this does not mean that the dynamics of the system are completely decoupled if $\cM \rar 0$, since $\cM$ quantifies the ``weight'' of the vector field with respect to the expansion rate, but there is additional dynamical information about the vector potential in $u^2$: if the vector field is very light in units of the expansion it could anyhow affect the expansion and the shear.

%%%%%%%%%%%%%%%%%%%%%%%%%%%%%%%%%%%%%%%%%%%%%%%%%%%%%%%%%%%%%%%%%%%
\section{Classifying the fixed points}\label{phase}
%%%%%%%%%%%%%%%%%%%%%%%%%%%%%%%%%%%%%%%%%%%%%%%%%%%%%%%%%%%%%%%%%%%

%%%%%%%%%%%%%%%%%%%%%%%%%%%%%%%%%%%%%%%%%%%%%%%%%%%%%%%%%%%%%%%%%%%
\subsection{Ultra-relativistic regime}

In the $\tg \rar 0$ case we obtain a very simple closed system:
\ba
(x^2)_{,\eN} & = & (2\en-3)x^2 -(1+4\Sa)u^2 -2\sqrt2 \cM u v \,, \\
u_{,\eN} & = & (\en+2\Sa-1)u + \sqrt2 \cM v \,, \\
v_{,\eN} & = & (\en-2\Sa-2)v \,, \\
\Sa_{,\eN} & = & (\en-3)\Sa +2v^2 \,, \\
\cM_{,\eN} & = & \en \cM \,.
\ea
Quite remarkably, there is no more reference to the three slopes $\lV$, $\lK$, and $\lM$: this means that the structure of the scalar functions $C(\phi)$, $D(\phi)$, and $M(\phi)$ do not influence the dynamics directly in this limit; nonetheless, the stability properties do depend on $\lK$ since it is this slope which tells us whether $\tg\rar0$ or not. The fixed points are in this case identical to the warped brane scenario: 1) de Sitter expansion with $y=\pm 1$ for any $\cM$ driven by the potential of the scalar field; 2) kinetic domination with $x=\pm 1$ (and everything else set to zero); 3) anisotropic solutions with $\Sigma = \pm 1$.  Thus, either there is no anisotropy, or the anisotropy is very large.

%%%%%%%%%%%%%%%%%%%%%%%%%%%%%%%%%%%%%%%%%%%%%%%%%%%%%%%%%%%%%%%%%%%
\subsection{Nonrelativistic regime}

Let us move on to the limit where $\tg \rar 1$.  It is clear at first glance that this case is much more interesting:
\ba
x_{,\eN} & = & (\en-3)x - \lM u^2 - \lV y^2 \,, \\
u_{,\eN} & = & (\en+2\Sa-1+\lM x)u + \cM v \,, \\
v_{,\eN} & = & (\en-2\Sa-2)v - \cM u \,, \\
\Sa_{,\eN} & = & (\en-3)\Sa -2u^2 + 2v^2 \,, \\
\cM_{,\eN} & = & (\en+\lM x) \cM \,.
\ea
In order to find the fixed points of the system we begin with $\cM\rar0$, i.e., we look at solutions for which the vector is light with respect to the expansion rate; this trivially solves the last equation. We were also able to obtain some fixed points with arbitrary $\cM$ but only for very special values of the coefficients $(\lV,\lM)$ which we list together with their $\cM=0$ relatives.

In studying the stability of the fixed points we follow the definitions of Lyapunov as in our earlier work on this subjet, see \cite{Koivisto:2014gia}:

Once again, in order to check whether $\tg=1$ is a stable solution we need to expand around this value: the coefficient of the linear term in $(1-\tg)\geq0$, whose sign we need to determine, is given by:
\be\label{tgstab}
(1-\tg)_{,\eN} = -2\lp 3 - \lK x + \lV y^2/x + \lM u^2/x \rp (1-\tg) \,.
\ee

There are five families of fixed points, which are described in turn below. For each fixed point, we first specify the solution, then write down the conditions for its existence and stability, and then briefly comment upon its potential applications in cosmology.

%%%%%%%%%
\subsubsection{\scshape{Isotropic scalar field domination}}

\ba
\begin{dcases}
  x = -\lV/3 \,,	& y = \sqrt{1-\lV^2/9} \,, \\
  u = 0 \,,		& v = 0 \,, \\
  \Sa = 0\,,		& \phantom{yo}\text{\small and}\,\, \en = \lV^2/3 \,.
\end{dcases}
\ea
\begin{center}
\begin{tabular}{|l|l|}
  \hline
  \multicolumn{2}{|c|}{$\cM=0$} \\ \hline
  Existence	& $|\lV|\leq3$ \\ \hline
  Stable	& $0<\lV<\sqrt6$ and $\lM>\lV$ \\
  		& $-\sqrt6<\lV<0$ and $\lM<\lV$ \\ \hline
  Semistable	& $\lM=\lV\,,\,\,|\lV|=3\,,\,\,|\lV|=\sqrt6\,,\,\,\lM=\lV-3/\lV$ \\ \hline
  \multicolumn{2}{c}{} \\[-8pt] \hline
  \multicolumn{2}{|c|}{$\cM\neq0$} \\ \hline
  Existence	& $\lM=\lV$ or $\lV=0$ \\ \hline
  Semistable	& $6|\cM|\leq|\lV^2-3|$ \\ \hline
  Lyapunov	& $6|\cM|>|\lV^2-3|$ \\ \hline
  \multicolumn{2}{c}{} \\[-8pt] \hline
  \multicolumn{2}{|c|}{$\tg\rar1$} \\ \hline
  \multicolumn{2}{|l|}{$\lK>0$ and either $\lV<0$ or $\lV>-\lK$} \\
  \multicolumn{2}{|l|}{$\lK<0$ and either $\lV>0$ or $\lV<-\lK$} \\
  \hline  
\end{tabular}
\end{center}

{$\boldsymbol \triangleright$ }{\it Comments --} This is a valid accelerating solution. Not the fixed point itself but its stability properties are affected by the vector field.

%%%%%%%%%
\subsubsection{\scshape{Kinetic driven anisotropic universe}}

\ba
\begin{dcases}
  x = x \,,		& y = 0 \,, \\
  u = 0 \,,		& v = 0 \,, \\
  \Sa = \sqrt{1-x^2}\,,	& \phantom{yo}\text{\small and}\,\, \en = 3 \,.
\end{dcases}
\ea

{$\boldsymbol \triangleright$ }{\it Comments --} This point with $\cM=0$ is not stable: there is always one eigenvalue which has positive real part.  Also the $\cM\neq0$ fixed point defined by $x=-3/\lM$ and $|\lM|\geq3$ are not stable.

%%%%%%%%%
\subsubsection{\scshape{Anisotropic scaling solution 1}}

\ba
\begin{dcases}
  x = -\frac{6\lV}{\lV^2+12} \,,	& y = \frac{3\sqrt{24-2\lV^2}}{\lV^2+12} \,, \\
  u = 0 \,,				& v = \pm\frac{\sqrt{(\lV^2-12)(18-3\lV^2)}}{\lV^2+12} \,, \\
  \Sa = \frac{2\lV^2-12}{\lV^2+12}\,,	& \phantom{yo}\text{\small and}\,\, \en = \frac{6\lV^2}{\lV^2+12} \,.
\end{dcases}
\ea
\begin{center}
\begin{tabular}{|l|l|}
  \hline
  \multicolumn{2}{|c|}{$\cM=0$} \\ \hline
  Existence	& $\sqrt6\leq|\lV|\leq2\sqrt3$ \\ \hline
  Stable	& $\sqrt6<\lV<2\sqrt3$ and $\lM>\lV$ \\
  		& $-2\sqrt3<\lV<-\sqrt6$ and $\lM<\lV$ \\ \hline
  Semistable	& $|\lV|=2\sqrt3\,,\,\,|\lV|=\sqrt6$ \\
		& $\sqrt6\leq|\lV|<6\sqrt{3/17}$ and either $\lM=\lV$ or $\lM=3(\lV/2-2/\lV)$ \\ \hline
  Lyapunov	& $6\sqrt{3/17}\leq|\lV|<2\sqrt3$ and either $\lM=\lV$ or $\lM=3(\lV/2-2/\lV)$\\ \hline
  \multicolumn{2}{c}{} \\[-8pt] \hline
  \multicolumn{2}{|c|}{$\cM\neq0$} \\ \hline
  Existence	& $\lM=\lV$ and either $|\lV|=2\sqrt3$ or $|\lV|=\sqrt6$ \\ \hline
  Semistable	& $|\lV|=\sqrt6$ and $|\cM|\leq1/2$ \\ \hline
  Lyapunov	& $|\lV|=\sqrt6$ and $|\cM|>1/2$ \\ \hline
  \multicolumn{2}{c}{} \\[-8pt] \hline
  \multicolumn{2}{|c|}{$\tg\rar1$} \\ \hline
  \multicolumn{2}{|l|}{$\lK>0$ and either $\lV<0$ or $\lV>-\lK$} \\
  \multicolumn{2}{|l|}{$\lK<0$ and either $\lV>0$ or $\lV<-\lK$} \\
  \hline  
\end{tabular}
\end{center}

{$\boldsymbol \triangleright$ }{\it Comments --}{ This solution is stable in some parameter range, however it cannot combine accelerating expansion with small shear since one always has $\Sa = -1 + \en/2$.}

%%%%%%%%%
\subsubsection{\scshape{Anisotropic scaling solution 2}}

\ba
\begin{dcases}
  x = \frac{3}{\lM-2\lV} \,,					& y = \frac12 \frac{\sqrt{6\lM^2-9\lM\lV+36}}{\lM-2\lV} \,, \\
  u = \pm\frac12 \frac{\sqrt{9\lV^2-6\lM\lV-36}}{\lM-2\lV} \,,	& v = \pm\frac12 \frac{\sqrt{6\lV^2-6\lM^2+3\lM\lV-36}}{\lM-2\lV} \,, \\
  \Sa = \frac12 \frac{\lV-2\lM}{\lM-2\lV}\,,			& \phantom{yo}\text{\small and}\,\, \en = \frac{3\lV}{2\lV-\lM} \,.
\end{dcases}
\ea
\begin{center}
\begin{tabular}{|l|rl|}
  \hline
  \multicolumn{3}{|c|}{$\cM=0$} \\ \hline
  Existence	& (i)	& \hspace*{-15pt} $|\lV| = 4\sqrt{6/17}$ and $|\lM| = \sqrt{6/17}$ \\
		& (ii)	& \hspace*{-15pt} $|\lV| = 2\sqrt3$ and either $|\lM| = 2\sqrt3$ or $|\lM|\leq\sqrt3$ \\
		& (iii)	& \hspace*{-15pt} $(3\lV+\sqrt{9\lV^2-96}) \leq 4\lM \leq (\lV+\sqrt{17\lV^2-96})$ \\
		& 	& \hspace*{-10pt} and either $\lV < -2\sqrt3$ or $4\sqrt{2/3} < |\lV| < 2\sqrt3$ \\
		& (iv)	& \hspace*{-15pt} $(\lV-\sqrt{17\lV^2-96}) \leq 4\lM \leq (\lV+\sqrt{17\lV^2-96})$ \\
		& 	& \hspace*{-10pt} and $4\sqrt{6/17} < |\lV| < 4\sqrt{2/3}$ \\
		& (v)	& \hspace*{-15pt} $(\lV-\sqrt{17\lV^2-96}) \leq 4\lM \leq (3\lV-\sqrt{9\lV^2-96})$ \\
		& 	& \hspace*{-10pt} and either $\lV > 2\sqrt3$ or $4\sqrt{2/3} < |\lV| < 2\sqrt3$ \\ \hline
  Semistable	& \multicolumn{2}{|l|}{$\lM=\lV$ with $|\lV|=2\sqrt3$} \\ \hline
  \multicolumn{3}{c}{} \\[-8pt] \hline
  \multicolumn{3}{|c|}{$\cM\neq0$} \\ \hline
  Existence	& \multicolumn{2}{|l|}{$\lM=\lV$ and $|\lV|=2\sqrt3$} \\
  \hline  
\end{tabular}
\end{center}

{$\boldsymbol \triangleright$ }{\it Comments --} This fixed point is unstable everywhere but for the very specific choice $\cM=0$, $\lM=\lV$ with $|\lV|=2\sqrt3$, for which $u=v=0$ and $\Sa=1/2$, see above.

%%%%%%%%%
\subsubsection{\scshape{Anisotropic scaling solution 3}}

\ba
\begin{dcases}
  x = -\frac{9\lV}{2\lV^2+12} \,,			& y = 0 \,, \\
  u = \pm \frac{3\sqrt{9-3\lV^2}}{2\lV^2+12} \,,	& v = \pm \frac{\sqrt{54+18\lV^2-12\lV^4}}{2\lV^2+12} \,, \\
  \Sa = \frac{4\lV^2-3}{2\lV^2+12}\,,			& \phantom{yo}\text{\small and}\,\, \en = \frac{6\lV^2+9}{\lV^2+6} \,.
\end{dcases}
\ea
\begin{center}
\begin{tabular}{|l|l|}
  \hline
  \multicolumn{2}{|c|}{$\cM=0$} \\ \hline
  Existence	& $\lM=2\lV$ and $|\lV|<\sqrt3$, or $|\lV|=\sqrt3$ \\ \hline
  \multicolumn{2}{c}{} \\[-8pt] \hline
  \multicolumn{2}{|c|}{$\cM\neq0$} \\ \hline
  Existence	& $|\lV|=\sqrt3$ and $\lM=2\lV$ \\
  \hline  
\end{tabular}
\end{center}

{$\boldsymbol \triangleright$ }{\it Comments --} Nowhere in its domain is this solution stable.

%%%%%%%%%
\subsubsection{\scshape{Anisotropic scaling solution 4}}

\ba
\begin{dcases}
  x = \frac{3(\lM-2\lV)}{3\lM^2-5\lM\lV+2\lV^2+12} \,,			& y = \frac{\sqrt{(3\lM^2-4\lM\lV+12)(3\lV^2-3\lM\lV+18)}}{3\lM^2-5\lM\lV+2\lV^2+12} \,, \\
  u = \pm \frac{\sqrt{(3\lM^2-4\lM\lV+12)(3\lM^2-3\lM\lV-9)}}{3\lM^2-5\lM\lV+2\lV^2+12} \,,	& v = 0 \,, \\
  \Sa = -\frac{2(\lV^2-\lM\lV-3)}{3\lM^2-5\lM\lV+2\lV^2+12}\,,		& \phantom{yo}\text{\small and}\,\, \text{ (i) } \en=3 \,, \text{\small or} \text{ (ii) } \en=\frac{9\lV^2}{10\lV^2+12} \,.
\end{dcases}
\ea
\begin{center}
\begin{tabular}{|l|rl|}
  \hline
  \multicolumn{3}{|c|}{$\cM=0$} \\ \hline
  Existence	& (i)	& \hspace*{-15pt} $\lV = 3(\lM/4+1/\lM)$ \\
		& (ii)	& \hspace*{-15pt} $\lV=-\lM$ and $|\lM|\geq\sqrt{3/2}$ \\ \hline
  \multicolumn{3}{c}{} \\[-8pt] \hline
  \multicolumn{3}{|c|}{$\cM\neq0$} \\ \hline
  Existence	& \multicolumn{2}{|l|}{$\lM=\lV$ and $|\lM|=2\sqrt3$} \\
  \hline  
\end{tabular}
\end{center}

{$\boldsymbol \triangleright$ }{\it Comments --} Once again, there is no stable domain.

%%%%%%%%%%%%%%%%%%%%%%%%%%%%%%%%%%%%%%%%%%%%%%%%%%%%%%%%%%%%%%%%%%%
\section{Numerical examples}\label{numerical}
%%%%%%%%%%%%%%%%%%%%%%%%%%%%%%%%%%%%%%%%%%%%%%%%%%%%%%%%%%%%%%%%%%%

Let us gain some intuition about some of these solutions by looking at the numerical evolution of the system.  The initial conditions are $x_0=y_0=1/4$ and $u_0=v_0=1/2$.

\begin{figure}
\centering
\includegraphics[width=0.35\textwidth]{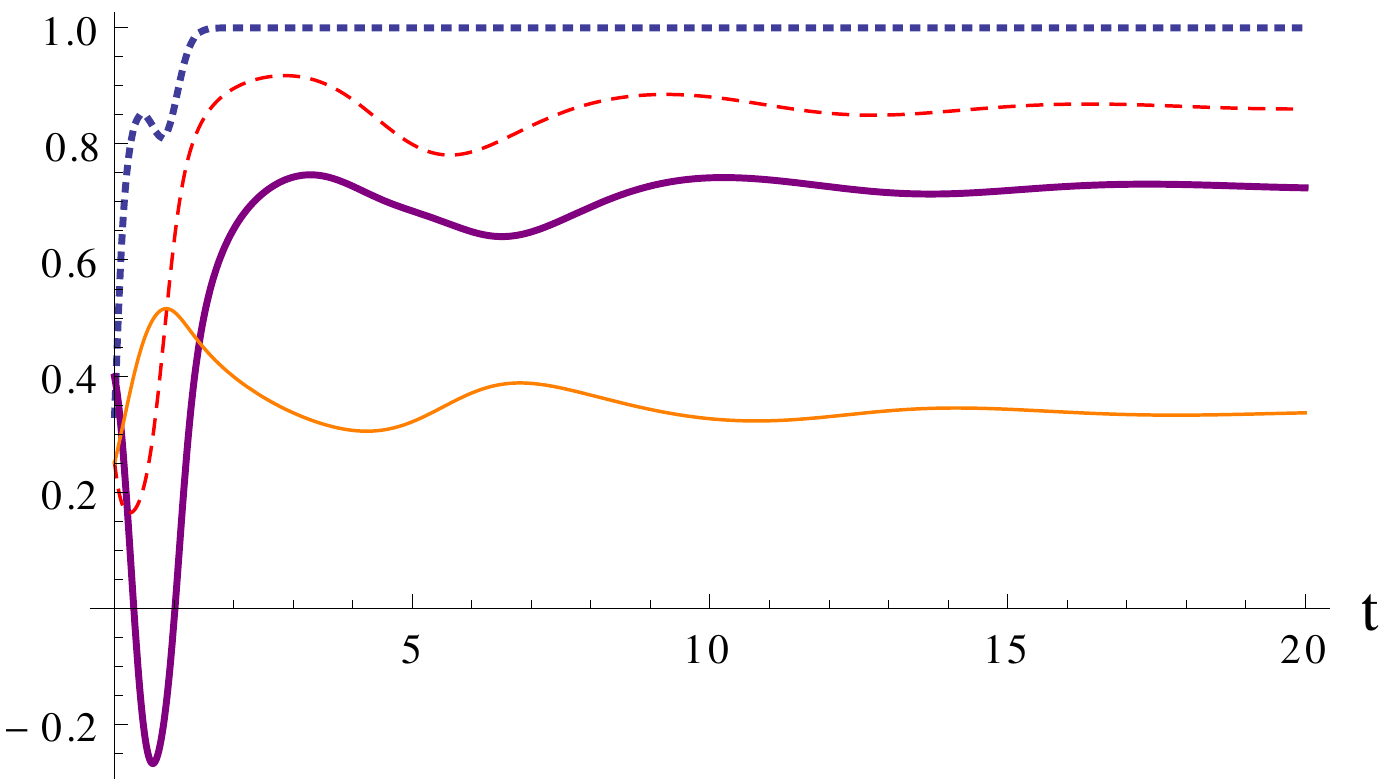}\hspace*{10pt}
\includegraphics[width=0.35\textwidth]{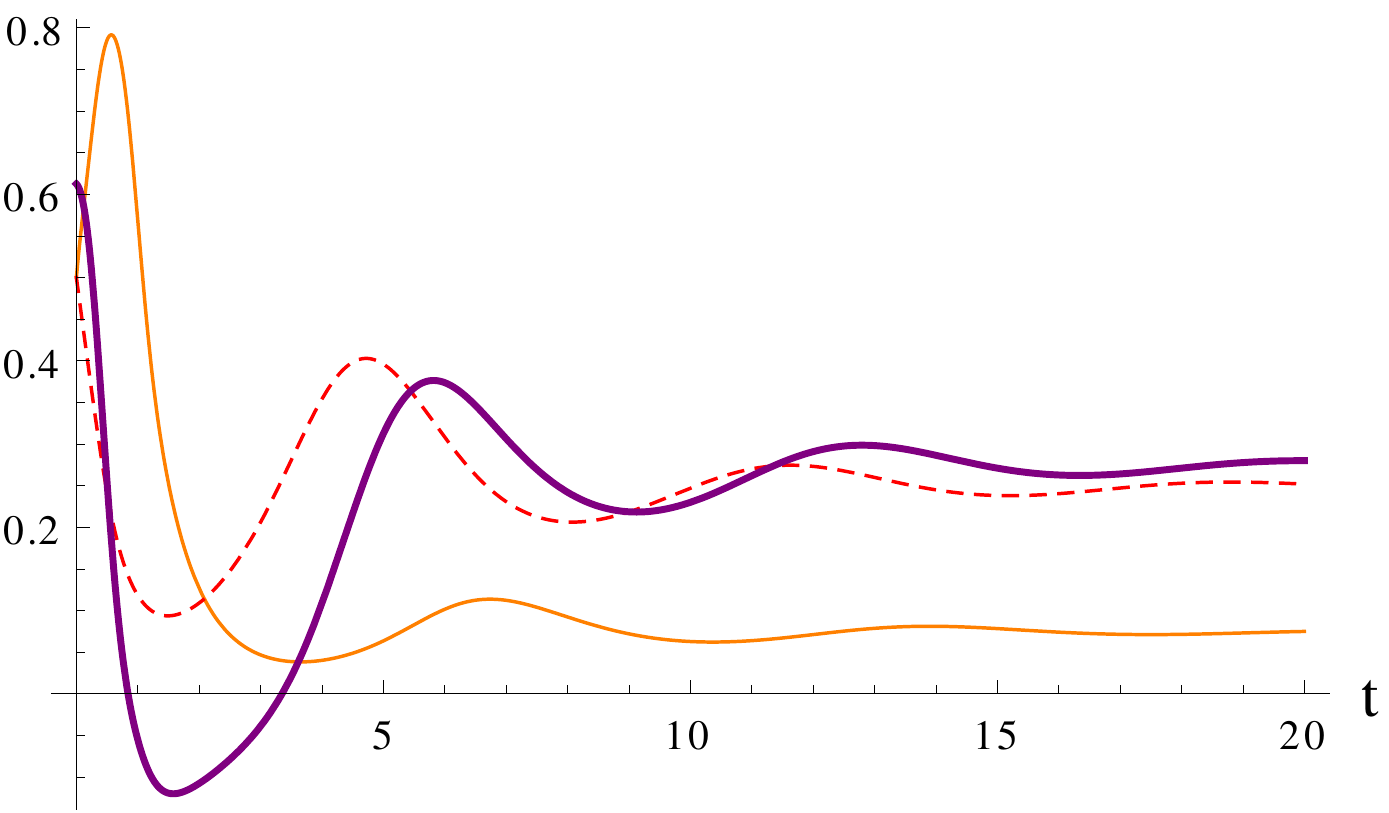}\hspace*{10pt}
\includegraphics[width=0.24\textwidth]{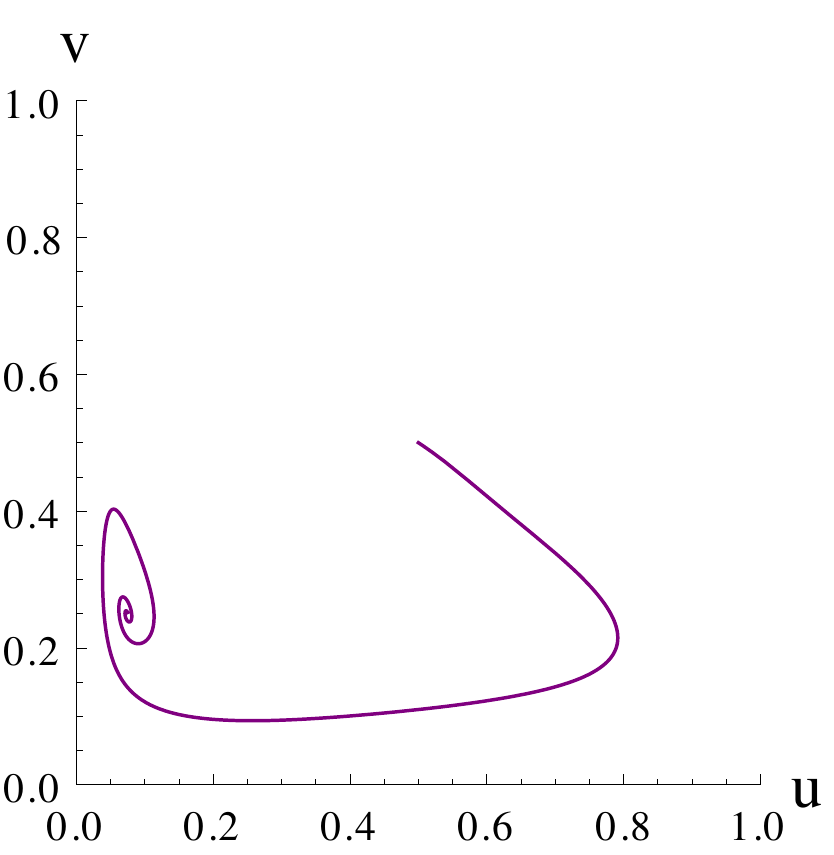}
\caption{The model $\lV=\lK=\lM=-3$, with $\tg_0=1/3$ and $\cM_0=1/10$.
{\bf Left panel}: The scalar field evolution. The dashed red line is $x$, the solid orange line is $y$. The $\tg$ is the thick dotted blue line and the thick purple line is the total equation of state.
{\bf Middle panel}: The vector field evolution. The dashed red line is $v$, the solid orange line is $v$. The shear $\Sa$ is the thick purple line.
{\bf Right panel}: The phase space evolution in the $(u,v)$ plane.}
\label{fig:model1}
\end{figure}

We begin with one of our attractor solutions, the Lyapunov stable point $\lV=\lK=\lM=-3$, which we follow for $\tg_0=1/3$ and $\cM_0=1/10$, see Figure \ref{fig:model1}.  Here $\tg$ quickly attains its non-relativistic attractor, while the vector and the scalar quantities all oscillate with decaying amplitudes: the limit for $t\rar\infty$ does not exist (due to the oscillations) but the point is nevertheless a Lyapunov attractor, as all quantities remain bounded, including the non-zero shear $\Sa$.

\begin{figure}
\centering
\includegraphics[width=0.35\textwidth]{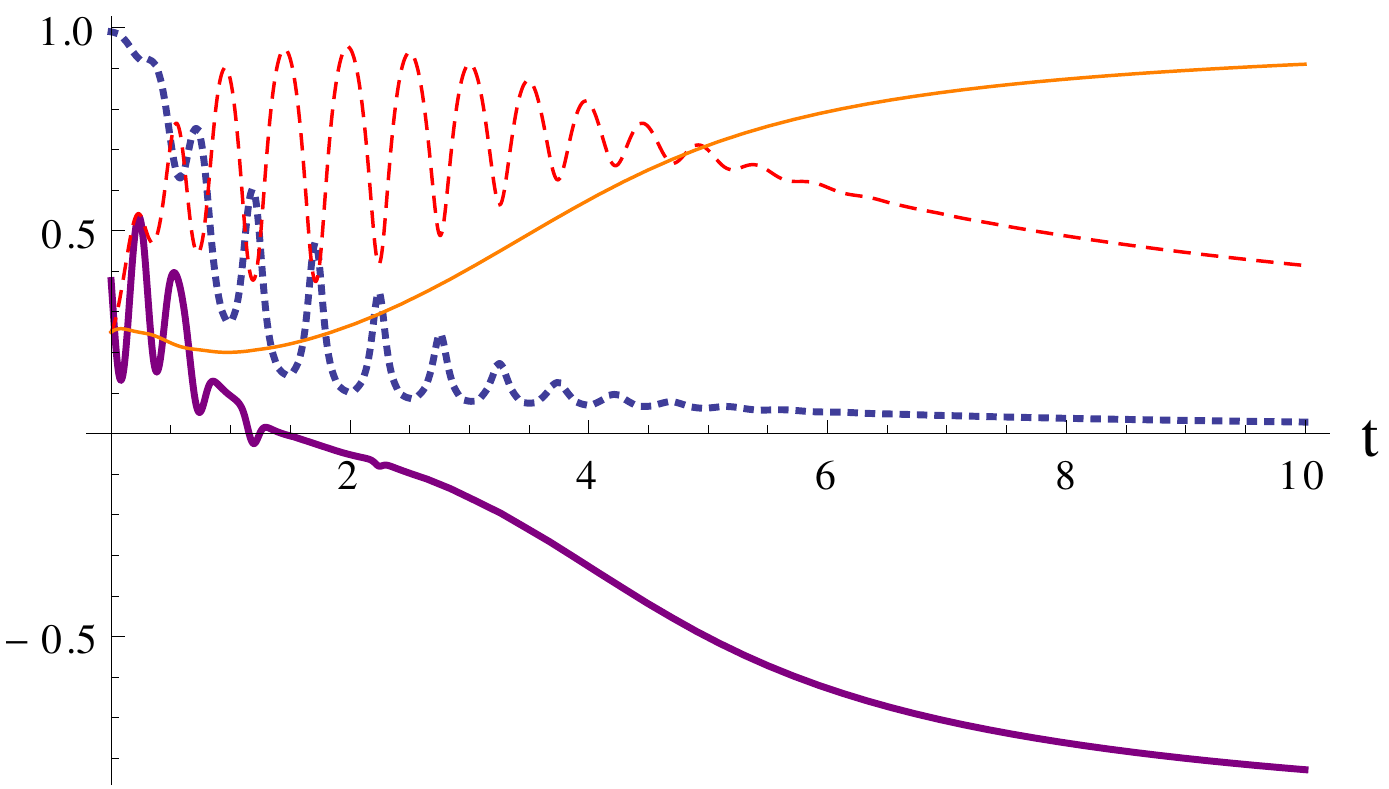}\hspace*{10pt}
\includegraphics[width=0.35\textwidth]{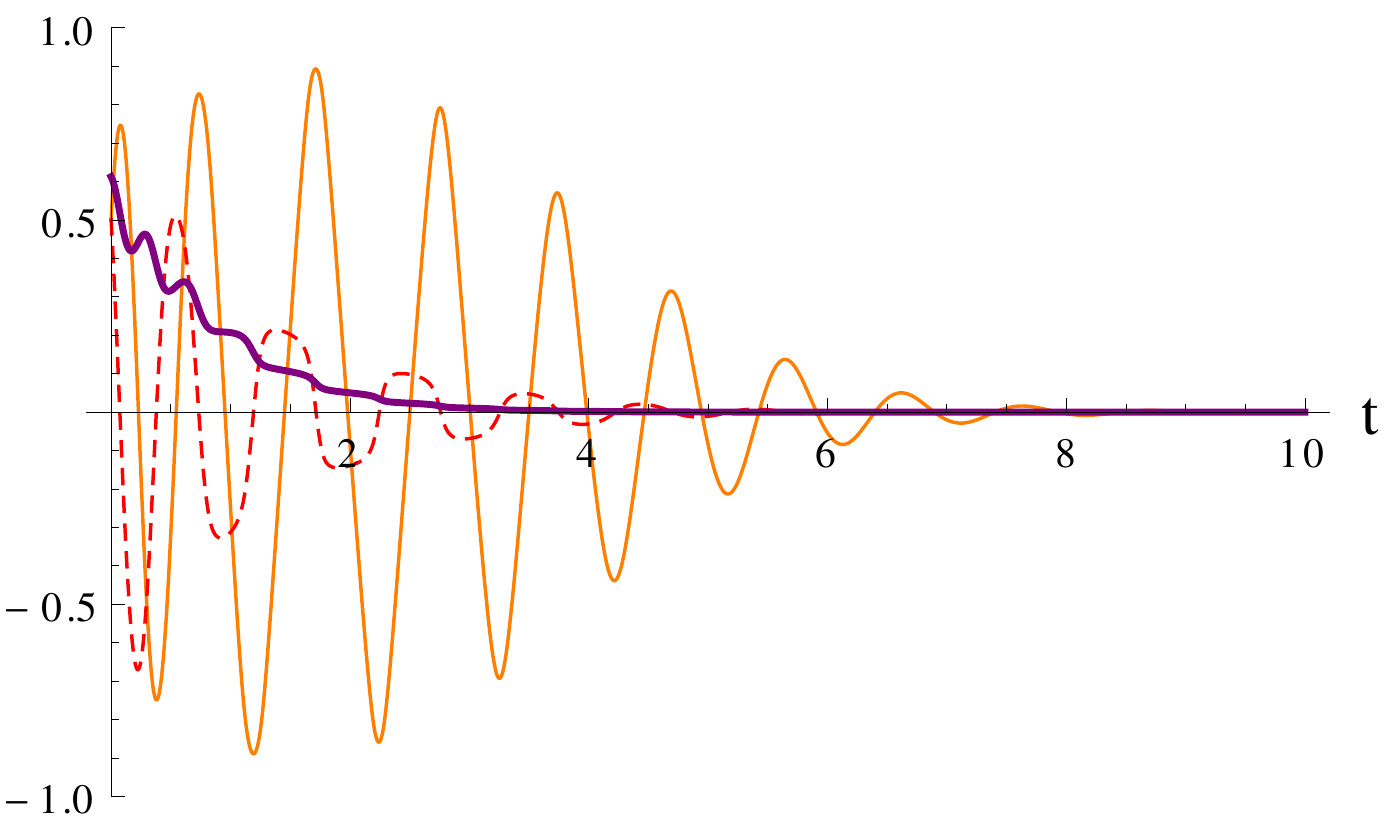}\hspace*{10pt}
\includegraphics[width=0.24\textwidth]{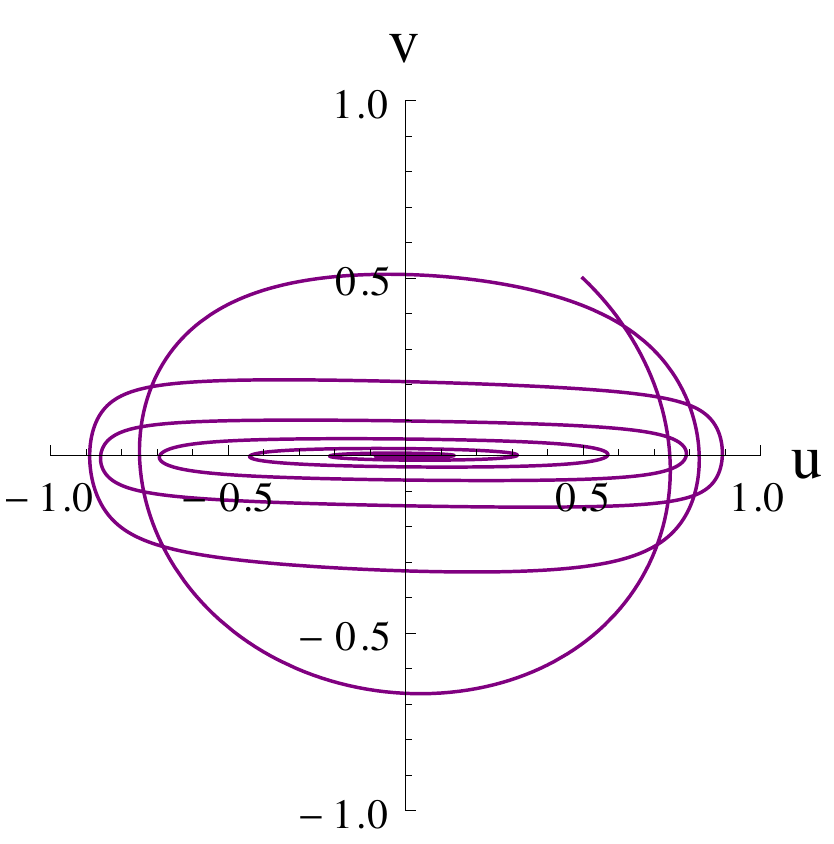}
\caption{The model $\lV=-5$, $\lK=8$, $\lM=-4$, with $\tg_0=99/100$ and $\cM_0=10$.
{\bf Left panel}: The scalar field evolution. The dashed red line is $x$, the solid orange line is $y$. The $\tg$ is the thick dotted blue line and the thick purple line is the total equation of state.
{\bf Middle panel}: The vector field evolution. The dashed red line is $v$, the solid orange line is $v$. The shear $\Sa$ is the thick purple line.
{\bf Right panel}: The phase space evolution in the $(u,v)$ plane.}
\label{fig:model2}
\end{figure}

In the following three examples we set $\lM=-\lK/2$ \footnote{This choice is reminiscent of a stringy interpretation of this model in terms of braneworlds.  If $h = D/C \sim \exp{(2\kappa\lambda\phi)}$, then $C \sim  1/D \sim \exp{(-\kappa\lambda\phi)}$.}.  In Figure \ref{fig:model2} we plot the model $\lV=-5$, $\lK=8$. We start from $\tg_0=99/100$ and a heavy $\cM_0=10$.  In this case, the asymptotic future attractor will instead be an ultra-relativistic ($\tg\rar0$) with contributions to the expansion rate from both the vector and the scalar field.  We see that the vector field and the scalar field oscillate before settling into the attractor: this is an effect of the large $\cM$, which will grow indefinitely.  Strictly speaking this is not a pure attractor because of the unbounded behaviour of $\cM$, but as we already mentioned, there is no dynamics attached to this variable.

\begin{figure}
\centering
\includegraphics[width=0.35\textwidth]{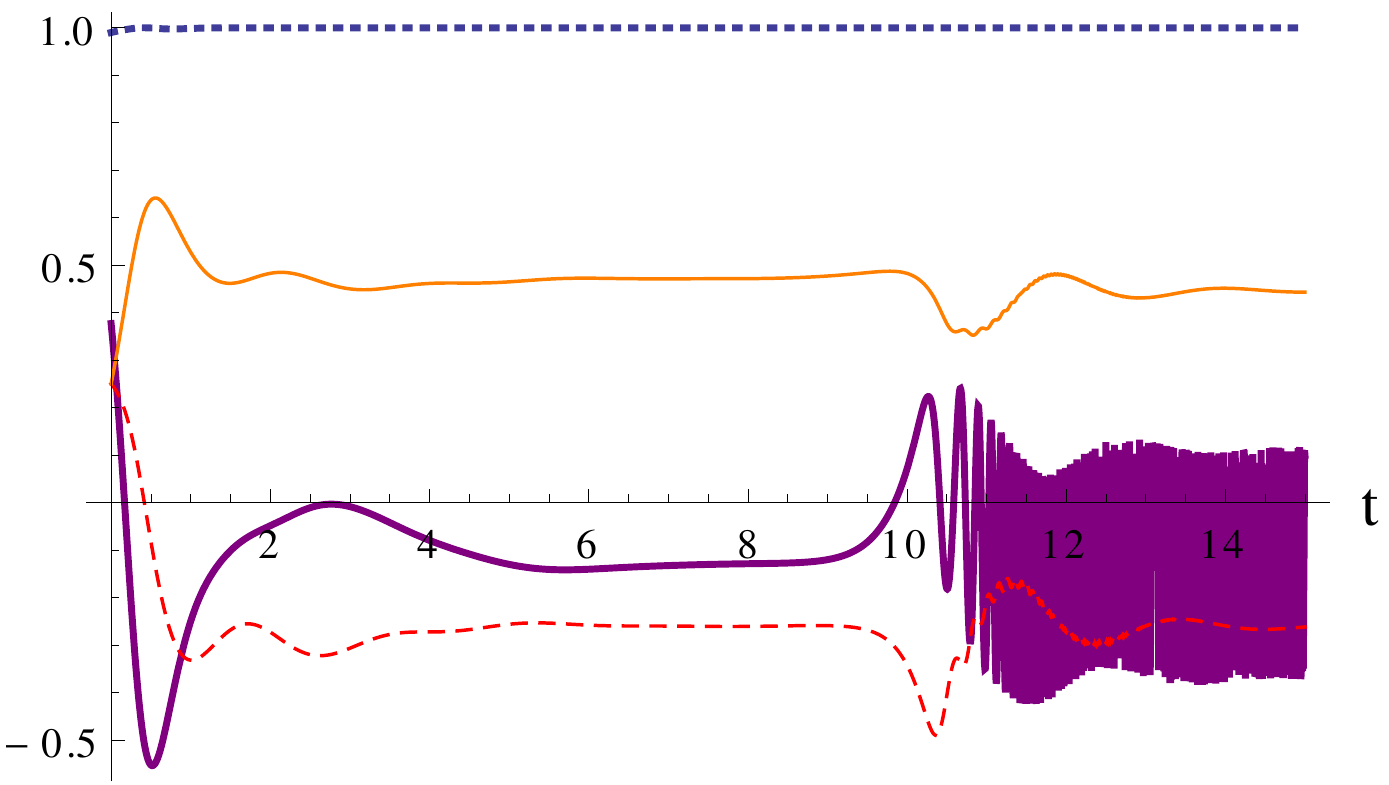}\hspace*{10pt}
\includegraphics[width=0.35\textwidth]{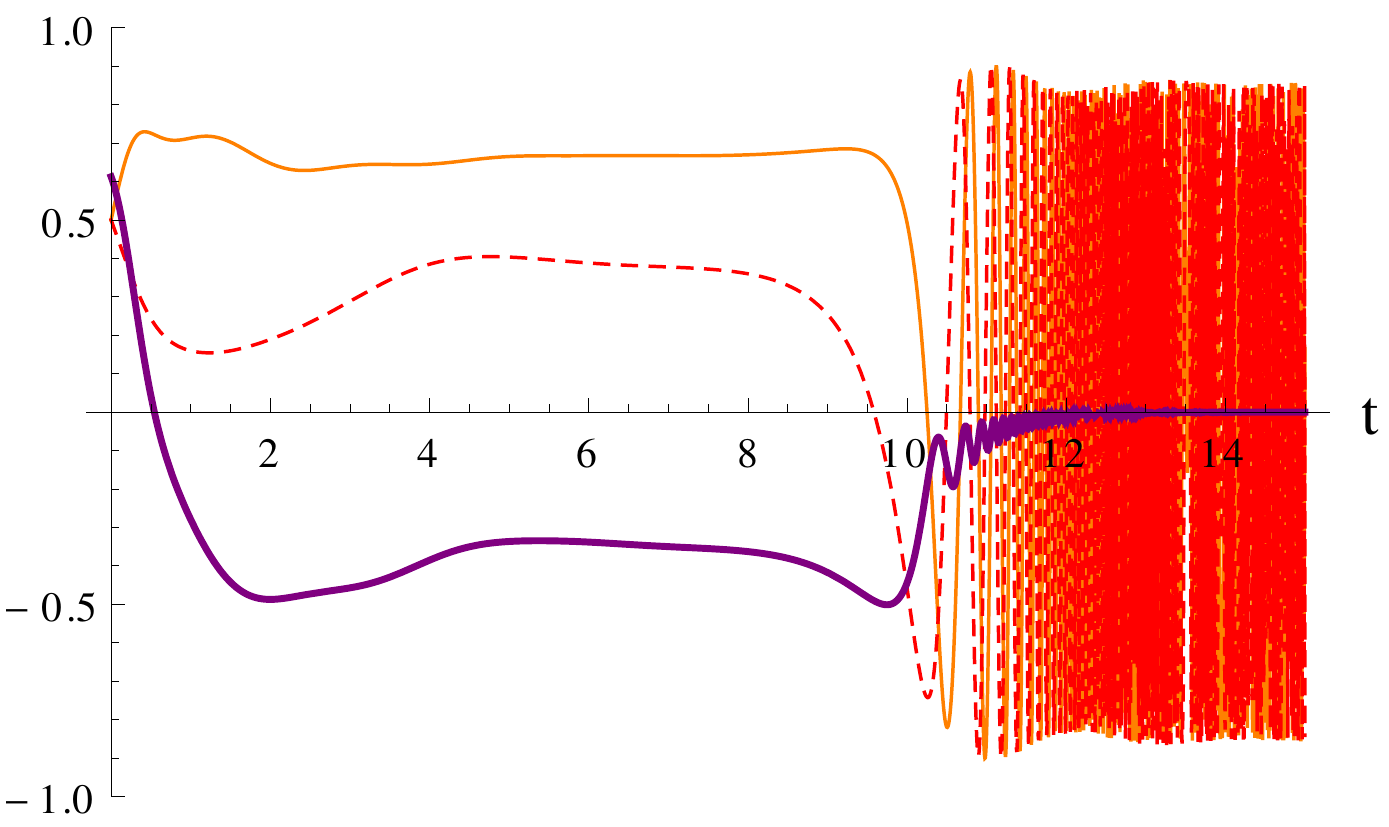}\hspace*{10pt}
\includegraphics[width=0.24\textwidth]{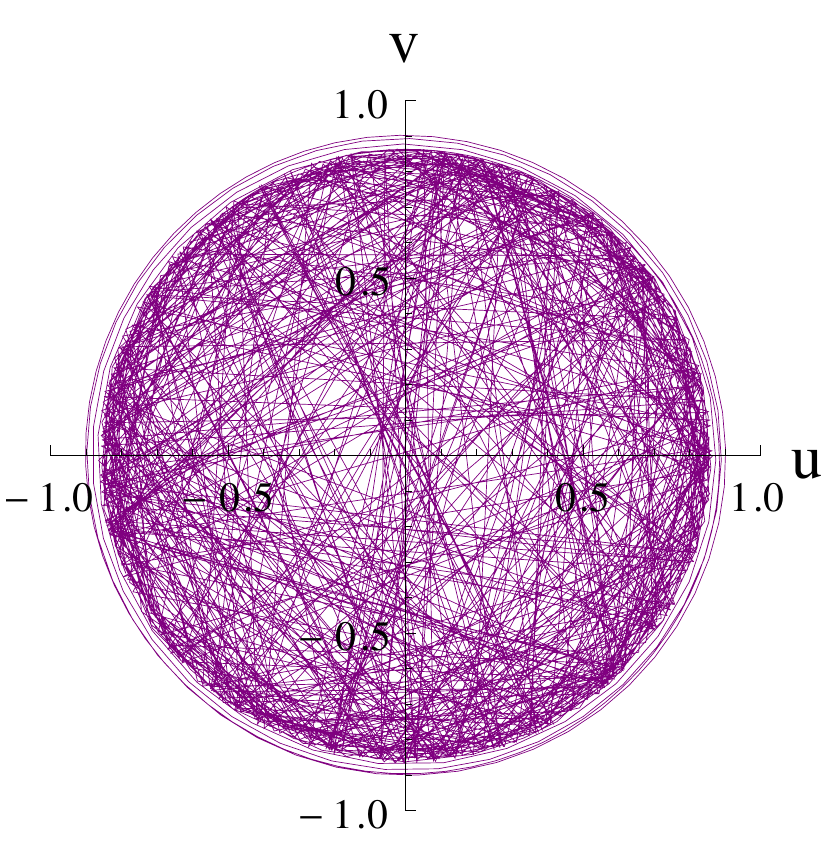}
\caption{The model $\lV=5$, $\lK=3$, $\lM=-3/2$, with $\tg_0=99/100$ and $\cM_0=10^{-7}$.
{\bf Left panel}: The scalar field evolution. The dashed red line is $x$, the solid orange line is $y$. The $\tg$ is the thick dotted blue line and the thick purple line is the total equation of state.
{\bf Middle panel}: The vector field evolution. The dashed red line is $v$, the solid orange line is $v$. The shear $\Sa$ is the thick purple line.
{\bf Right panel}: The phase space evolution in the $(u,v)$ plane.}
\label{fig:model3}
\end{figure}

In Figure \ref{fig:model3} we show the case of $\lV=5$, $\lK=3$.  Here the asymptotic future attractor will again be a non-relativistic ($\tg\rar1$) solution.  The vector will be strongly oscillating, so that neither $u$ nor $v$ are constants. However, their total contribution to the expansion rate, $u^2+v^2$, will be approximately a constant, and will scale with the scalar field contributions $x^2$ and $y^2$ with on average, a nearly dust-like equation of state.  This is an interesting solution as it shows that a spatial vector field is compatible with the FLRW symmetries, and can mimic a matter-like component.  Note that the shear decays in the evolution even though the vector field contribution does not: this is line with the isotropy theorem for cosmic vector fields put forward by in \cite{Cembranos:2012kk}.

\begin{figure}
\centering
\includegraphics[width=0.35\textwidth]{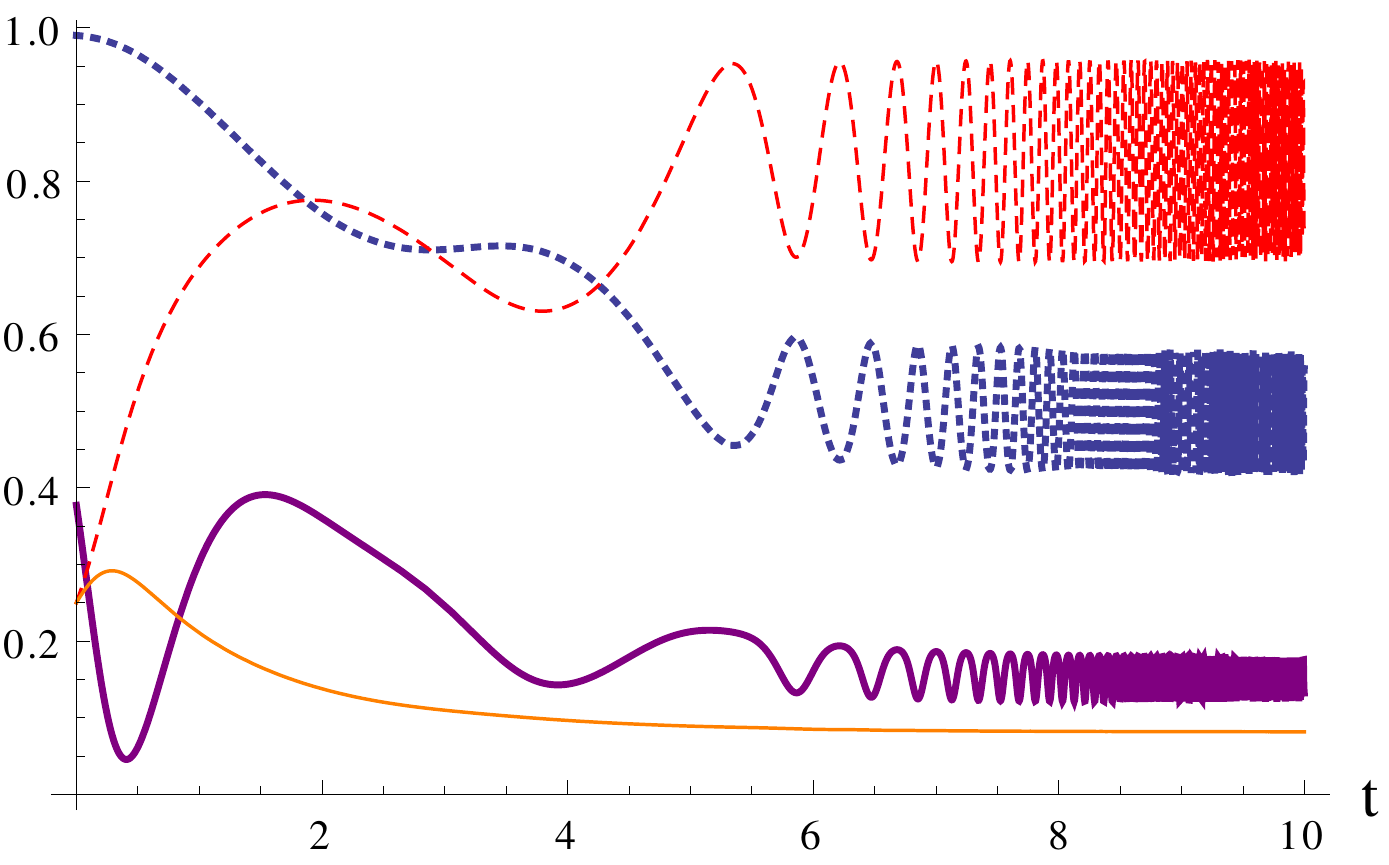}\hspace*{10pt}
\includegraphics[width=0.35\textwidth]{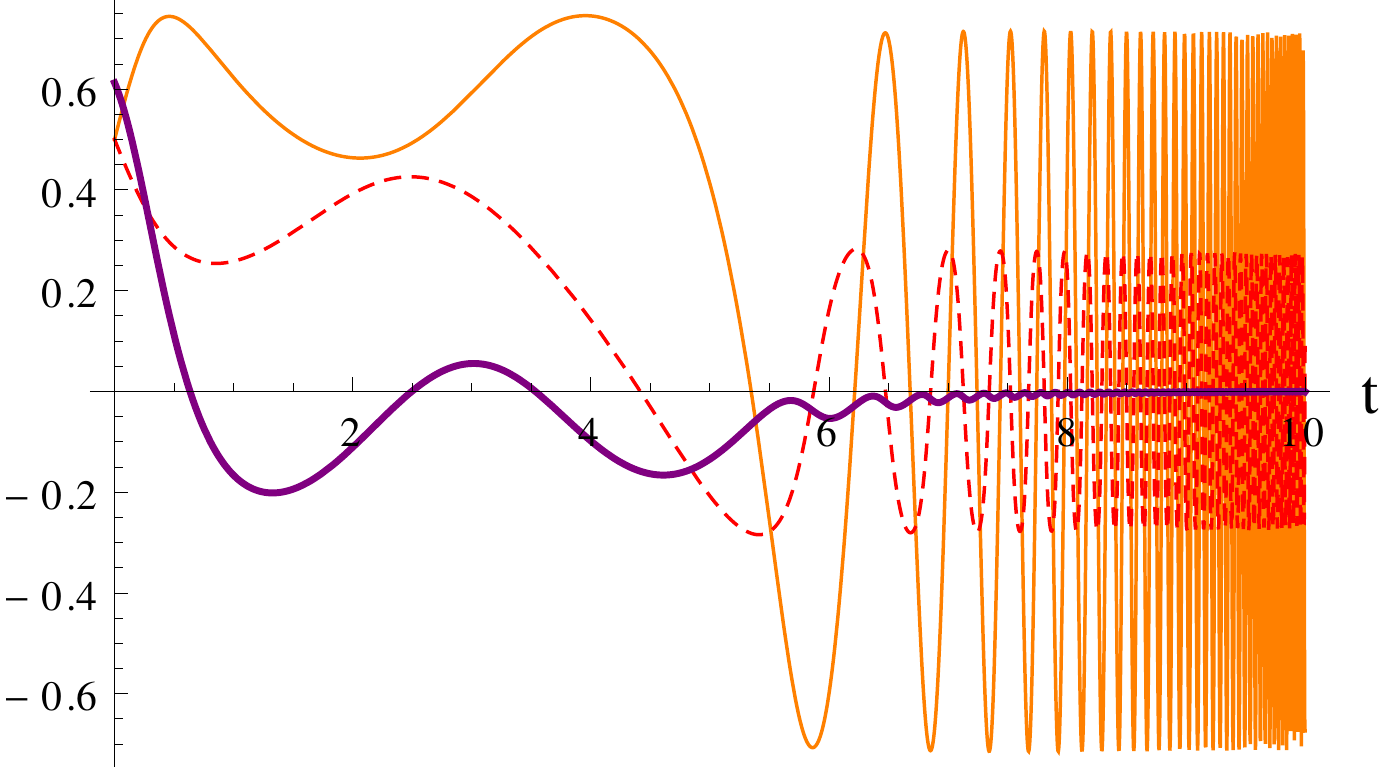}\hspace*{10pt}
\includegraphics[width=0.24\textwidth]{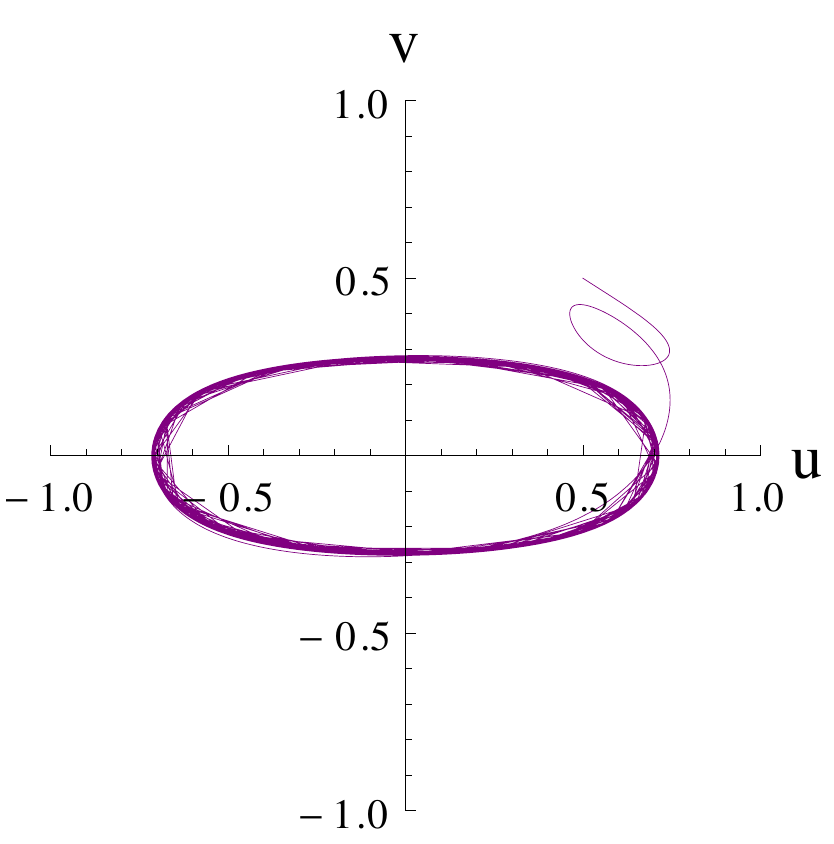}
\caption{The model $\lV=-\lK=2\lM=-4$, with $\tg_0=99/100$ and $\cM_0=1/30$.
{\bf Left panel}: The scalar field evolution. The dashed red line is $x$, the solid orange line is $y$. The $\tg$ is the thick dotted blue line and the thick purple line is the total equation of state.
{\bf Middle panel}: The vector field evolution. The dashed red line is $v$, the solid orange line is $v$. The shear $\Sa$ is the thick purple line.
{\bf Right panel}: The phase space evolution in the $(u,v)$ plane.}
\label{fig:model4}
\end{figure}

Finally, in Figure \ref{fig:model4} we plot the special case $\lV=-\lK=4$, with again $\tg_0=99/100$ but now non-negligible $\cM_0=1/30$.  When the $\lK$ and $\lV$ have equal magnitude but the opposite sign, fixed points with finite $\tg$ appear for the scalar field when it is dominated by its kinetic term.  Similar solutions for the DBI scalar were reported in the absence of the vector \cite{Copeland:2010jt}.  Once again notice how the overall shear decays, FLRW is recovered, albeit the contributions from the vector oscillate strongly and are not tamed.  The equation of state also strongly oscillates, and its final value depends ultimately on the choice of $\lV$.

%%%%%%%%%%%%%%%%%%%%%%%%%%%%%%%%%%%%%%%%%%%%%%%%%%%%%%%%%%%%%%%%%%%
\section{Conclusions}\label{conclusions}
%%%%%%%%%%%%%%%%%%%%%%%%%%%%%%%%%%%%%%%%%%%%%%%%%%%%%%%%%%%%%%%%%%%

In this work we studied the cosmological dynamics of a particular vector-tensor model of gravity which is a simple modification and extension of the simplest disformal vector-tensor theory, which arises in presense of a double boost factor for the vector sector.  Because of the presence of a single vector field, which can develop a non-trivial vacuum expectation value, we worked with an anisotropic Bianchi I spacetime, which includes anisotropic shear.  We were able to reduce the system to a first-order one which can be studied with usual dynamical systems techniques.  The equations of motion follow the evolution of a massive Abelian vector field, a generic real scalar field, and gravity, all of which were parametrised in such a way that the relatively cumbersome system becomes fairly manageable.  We did obtain a number of new fixed points, whose existence and stability properties we then studied systematically.  For the sake of concreteness, we picked exponential forms for both the scalar and vector potential and their coupling, that is, in first approximation, we kept the slopes $\lV$, $\lM$, and $\lK$ as constant.

The outcome of our analysis does share several aspects with the single-boost warped brane case we investigated in our previous work, in the sense that, also with the modification presented in this work, the shear, which could in principle be supported by anisotropic stress of the spatial vector field, decays and the metric isotropises towards flat FLRW.  Compared to previous work however we were able to obtain even more anisotropic fixed points, including scaling solutions; they however still seem not to be of immediate cosmological relevance because either they are not attractors, or it is not possible to have simultaneously a slowly rolling vacuum energy driving the expansion, and small shear.  Curiously, one can show that for all the fixed points we obtained, there exists an approximate relation $\en\sim\Sa+{\cal O}(1)$: this means that either $\en\ll1$, or $\Sa\ll1$, but not both.

This however does not mean that there is no cosmological application for this theory.  Indeed, as illustrated in our numerical examples, even in the absence of shear, the single vector can be compatible with the FLRW symmetries due to its very rapid oscillations, as predicated in~\cite{Cembranos:2012kk}.  Once the vector oscillated very rapidly its equation of state can be, on average, like that of dust, which in turn implies that such models can provide a disformally coupled alternative to dark matter.

%%%%%%%%%%%%%%%%%%%%%%%%%%%%%%%%%%%%%%%%%%%%%%%%%%%%%%%%%%%%%%%%%%%
\subsection*{Acknowledgements}

We would like to thank Danielle Wills for useful discussions.
FU is supported by IISN project No.~4.4502.13 and Belgian Science Policy under IAP VII/37.

%%%%%%%%%%%%%%%%%%%%%%%%%%%%%%%%%%%%%%%%%%%%%%%%%%%%%%%%%%%%%%%%%%%
%%%%%%%%%%%%%%%%%%%%%%%%%%%%%%%%%%%%%%%%%%%%%%%%%%%%%%%%%%%%%%%%%%%

\bibliography{Drefs}

\end{document}